# Cache Hierarchy Optimization

Leonid Yavits, Amir Morad, Ran Ginosar

Department of Electrical Engineering, Technion - Israel Institute of Technology, Haifa, Israel

**Abstract**— Power consumption, off-chip memory bandwidth, chip area and Network on Chip (NoC) capacity are among main chip resources limiting the scalability of Chip Multiprocessors (CMP). A closed form analytical solution for optimizing the CMP cache hierarchy and optimally allocating area among hierarchy levels under such constrained resources is developed. The optimization framework is extended by incorporating the impact of data sharing on cache miss rate. An analytical model for cache access time as a function of cache size is proposed and verified using CACTI simulation.

——————————— ◆ ———————————

## 1 INTRODUCTION

The advancement of CMP architectures has been very significant in recent years. Despite the progress of silicon technology, chip resources including power budget, off-chip memory bandwidth, chip area and NoC capacity remain limited [10]. Partitioning of the chip real-estate among various CMP components, most importantly between cores and cache, but also inside the cache, to maximize CMP performance under these constraints remains a critical dilemma for the computer architect. In this work we focus on cache and propose the optimization of the area allocation across hierarchy levels, under constrained power, off-chip memory bandwidth, chip area and NoC bandwidth.

Area allocation among CMP components has been extensively studied. Alameldeen [1] used analytical modeling to study the trade-off between the number of cores and cache size. Huh *et al.* [8] studied area and performance trade-offs. Oh *et al.* [13] developed analytical models of various cache organizations. Cassidy and Andreou [2] introduced a closed form solution to optimally allocate constrained area between core and cache in a symmetric CMP. Krishna *et al.* [3] researched the effects of data sharing in multithreaded applications on optimal area allocation between cores and cache.

Those studies mainly focus on division of area resources between cores and cache. In our work, we aspire to provide the architect with an analytical framework that yields the optimal hierarchy (the optimal number of private and shared hierarchy levels) and the optimal area allocation among hierarchy levels. Since average memory access delay is an additive component of the overall CMP CPI [3][13], our optimization results can be incorporated into a larger cores vs. cache optimization framework.

Another aspect of our work is cache access time modeling. Most existing studies assume that cache access time is known *a priori* and is constant, even though a very wide range of cache sizes is considered by the optimization process (for example, from 64KByte to 4MByte in [2]). In this work we propose an analytical model for varying cache access time as a function of cache size (following [2], we use the terms *cache area* and *cache size* interchangeably). This approach, in conjunction with modeling NoC delay as function of cache size, leads to a more realistic cache access time model.

In addition, we extend the optimization framework by modeling the effect of data sharing and remote data access (i.e. access to the data originated and processed in other cores) on the miss rate of the private cache.

The CACTI simulator [12] provides the architect an efficient tool for cache access time and area estimation, including bank, line and aspect ratio optimization. It does not suggest however the optimal cache hierarchy and/or area allocation among hierarchies. In order to find an acceptable (though potentially suboptimal) configuration, the architect may need many trial runs of CACTI. Our optimization framework can be incorporated into the CACTI simulator to offer an optimized cache hierarchy in addition to the optimized single-level cache organization.

The rest of this paper is organized as follows. Section 2 presents the analytical model for cache access time and its verification using CACTI simulations. Section 3 presents the optimization framework. Section 4 offers conclusions.

## 2 ACCESS TIME AS A FUNCTION OF CACHE SIZE

Exploring the circuit level cache model detailed in [12] and [14] while varying the Block Size, Associativity and Number of Sets, we find that the access time of the $i^{th}$ level cache can be approximated by power-law model:

$$t_i(A_i) = \tau \cdot (A_i/\alpha)^\beta \quad (1)$$

where $A_i$ is the size of the $i^{th}$ level cache and is equal to Block Size × Associativity × Number of Sets; $\tau$ and $\alpha$ are the access time and the size of a baseline cache; the $\beta$ exponent is found by fitting the power law (1) curve to the cache access time data, either received by exploring circuit level models or generated by CACTI. It is mainly affected by the technology node, as shown in Fig. 1(a).

Two models for the hit latency as a function of cache size are compared here: $D_{CA}$, based on (36) in [2], assuming constant access time $t_1 = \chi$, is presented in (2) and our model $D_{YMG}$, using (1), is shown in (3):

$$D_{CA} = (1 - m_1)t_1 = \left(1 - \mu/\sqrt{A_1/\alpha}\right) \cdot \chi \quad (2)$$

$$D_{YMG} = (1 - m_1)t_1 = \left(1 - \mu/\sqrt{A_1/\alpha}\right) \cdot \tau \cdot (A_i/\alpha)^\beta \quad (3)$$

where $m_i$ is the miss rate of the $i^{th}$ level cache and $\mu$ is the miss rate of a baseline cache. In the constant access time model $D_{CA}$, the average memory delay is monotonically decreasing with the cache size (since miss rate decreases with cache size). In contrast, the $D_{YMG}$ model exhibits an interim minimum point and increasing the cache size beyond that point leads to longer average memory delay.

Some previous studies lay out the framework for optimal area allocation between cores and cache. We find that the optimal design point is substantially different when cache access time is modeled as the function of cache size. This observation is reflected in Fig. 1(b), which presents the cache per core area (vs. the number of cores), optimized by the constant access time based model developed by Cassidy and Andreou (red chart), and compared to empirical CMP data (scattered black dots) [2]. Our model (based on (16) in Section 3 and shown as the blue chart) is similar to the Cassidy and Andreou model for smaller scale CMPs (up to 64 cores). For larger CMPs (80+ cores), Cassidy and Andreou model indicates that those CMPs are undercapitalized in terms of cache per core, while our model suggests that cache and core areas are balanced rather reasonably.

The cache access times based on our power law model (1) vs. cache access time simulated for 45nm by CACTI 6 [11] and for 25nm by 3D CACTI [17] are shown in Fig. 1(c) and (d). The charts demonstrate that the power-law (1) approximates CACTI simulations (and their underlying circuit level models) over a wide range of cache sizes (from 4Kbytes to 16Mbytes) to within 5%. The constant cache access time assumption of [2] is indicated as a horizontal line in Fig. 1(c).

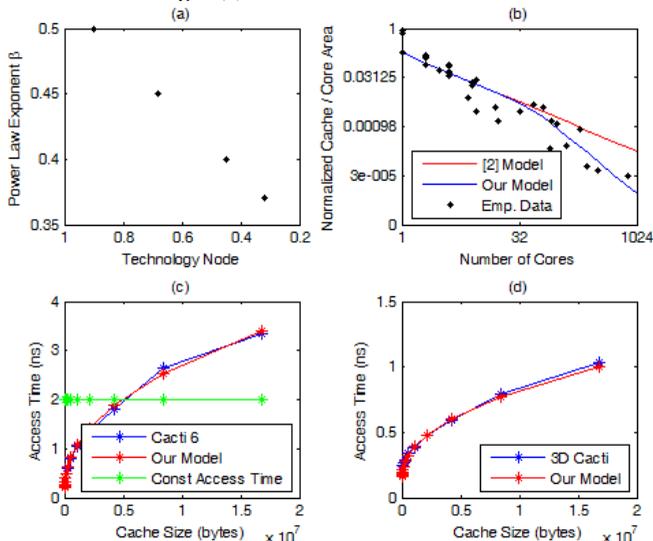

Fig. 1. (a) Power law exponent $\beta$ vs. technology node; (b) Normalized Cache per Core Area vs. Number of Cores; (c) Cache access time vs. Cache size (CACTI 6); (d) Cache access time vs. Cache size (3D CACTI)

## 3 OPTIMIZING CACHE HIERARCHY

In this work, we focus on three typical cache configurations: single private level, two-level (one private + one shared) and three level (two private + one shared) cache. The framework can be easily extended to any number of private, shared or hybrid levels. The access times of each private level $i = 1 \div 3$ and each shared level $j = 2 \div 3$ are:

$$t_i = \tau \cdot \left(A_i/\alpha\right)^\beta ; t_j = d_{NoC} + \tau \cdot \left(A_j/n\alpha\right)^\beta \quad (4)$$

where $n$ is the number of shared cache clients; NoC delay $d_{NoC}$ is a sum of transfer delay $d_t$, blocking delay $d_b$ and queuing (congestion) delay $d_c$. We adopt the analytical models of $d_b$ and $d_c$ proposed by [15] and [16]; $d_b$ and $d_c$ depend on a variety of parameters including the shared cache access rate $M_S$, the network capacity, the number of cores etc.. Those parameters except for $M_S$ are not part of our optimization framework. Therefore we model both $d_b$ and $d_c$ as function of $M_S$, assuming the rest of parameters are constant. Transfer delay $d_t$ is $O(\sqrt{n})$, assuming 2-D mesh NoC [7].

The average memory delays $D_1$, $D_{12}$ and $D_{123}$ for the above three configurations can be written as follows:

$$\begin{aligned} D_1 &= (1 - m_1)t_1 + m_1(d_D + d_Q); \\ m_1 &= \mu_n + (1 - \mu_n)\mu/\sqrt{A_1/\alpha}; \quad M_S = M_D = m_1 \end{aligned} \quad (5)$$

$$\begin{aligned} D_{12} &= (1 - m_1)t_1 + m_1(1 - m_2)t_2 + m_1 m_2(d_D + d_Q); \quad m_1 = \mu_n + (1 - \mu_n)\mu/\sqrt{A_1/\alpha}; \quad m_2 = \\ &\mu E_n/\sqrt{(A_2 - A_1)/(n\alpha)}; M_S = m_1; M_D = m_1 m_2 \end{aligned} \quad (6)$$

$$\begin{aligned} D_{123} &= (1 - m_1)t_1 + m_1(1 - m_2)t_2 + m_1 m_2(1 - m_3)t_3 + m_1 m_2 m_3(d_D + d_Q); \\ m_1 &= \mu_n + (1 - \mu_n)\mu/\sqrt{A_1/\alpha}; \\ m_2 &= \mu_n + (1 - \mu_n)\mu/\sqrt{(A_2 - A_1)/\alpha}; \\ m_3 &= \mu E_n/\sqrt{(A_3 - A_2)/(n\alpha)}; \\ M_S &= m_1 m_2; \quad M_D = m_1 m_2 m_3 \end{aligned} \quad (7)$$

where $M_D$ is the rate of access to off-chip DRAM, $d_D$ is the DRAM access penalty, $d_Q$ is the interconnect queuing delay, $E_n$ is the data sharing factor [3], and $\mu_n$ is the compulsory miss component, which reflects access to data originated in remote (rather than in local) core [9]; $\mu_n$ does not depend on the size of the local cache. DRAM interconnect queuing delay $d_Q$ can be presented as a function of the rate of access to off-chip DRAM $M_D$, the off-chip memory bandwidth, the number of cores etc. [3]. Those parameters except for $M_D$ are not part of our optimization framework. Hence we model $d_Q$ as a function of $M_D$, assuming the rest of the parameters are constant. Lastly, (6) and (7) assume inclusive cache and can be easily modified to support non-inclusive cache.

The objective function representing the average memory delay, yielded by the best of three possible configurations, under variety of constraint resources is obtained using KKT multipliers similarly to [4]:

$$D = min\{D_{123}, D_{12}, D_1\} + \sum \lambda_j \cdot [g_j(A_1, A_2, A_3) - L_j] \quad (8)$$

where $g()$ is the constraint function, $L$ is the resource limit and $\lambda$ is the KKT multiplier.

It has been suggested that the power consumption of cache scales as the square root of its area [2], therefore the power constraint can be written as follows:

$$g_1(A_1, A_2, A_3) = \sum_{j=1}^{3} \rho\sqrt{A_j/\alpha} \leq P_{max} \quad (9)$$

where $\rho$ is the power consumption of a baseline cache and $P_{max}$ is the maximum power available to cache. Note, we do not consider the power consumption of the NoC since it is common to all three configurations.

The way to restrict the off-chip memory traffic in our optimization framework is by limiting the rate of access to off-chip DRAM $M_D$:

$$g_2(A_1, A_2, A_3) = M_D(A_1, A_2, A_3) \leq M_{D\,max} \quad (10)$$

where $M_{D\,max}$ is the maximum off-chip DRAM bandwidth. The area constraint can be presented as the sum of areas of all cache levels, as follows:

$$g_3(A_1, A_2, A_3) = \sum_{j=1}^{3} A_j \leq A_{max} \quad (11)$$

where $A_{max}$ is the maximum area available to cache. Finally, we can restrict the NoC traffic by limiting the rate of access to the shared cache $M_S$:

$$g_4(A_1, A_2, A_3) = M_S(A_1, A_2, A_3) \leq M_{S\,max} \quad (12)$$

where $M_{S\,max}$ is the maximum capacity of the NoC. The unconstrained objective function $min\{D_{123}, D_{12}, D_1\}$ can be presented in a differentiable form similarly to [4]:

$$\begin{aligned} min\{D_{123}, D_{12}, D_1\} &= min\{min\{D_{12}, D_1\}, D_{123}\} \\ &= [D_1 H_1 + D_{12}(1 - H_1)] H_2 \\ &\quad + D_{123}(1 - H_2) \end{aligned} \quad (13)$$

where $H$ is the step function:

$$H_1 = \begin{cases} 1, D_1 < D_{12} \\ 0, D_1 > D_{12} \end{cases}; \quad H_2 = \begin{cases} 1, min\{D_{12}, D_1\} < D_{123} \\ 0, min\{D_{12}, D_1\} > D_{123} \end{cases} \quad (14)$$

The partial derivatives of $H_1$ and $H_2$ with respect to $A_i$ are zero except for those $A_1, A_2$ and $A_3$ where $H_1, H_2$ and their derivatives are not defined, that is, when any of the following equalities holds:

$$D_1 = D_{12}; \; D_1 = D_{123},; \; D_{12} = D_{123} \quad (15)$$

However, these equality points are of no consequence for the optimization (since the decision there can go either way yielding the same delay), and therefore can be omitted. Consequently, substituting various constraint functions (or their combinations) into (8) and differentiating the constrained objective function $D$ with respect to $A_j$ and $\lambda_j$, we can find the optimal hierarchy and optimal area allocation:

$$[\partial D/\partial A_j \; \partial D/\partial \lambda_j] = 0 \quad \forall j = 1 \div 3, \quad (16)$$

The system of equations (16) can be solved numerically, using assumptions similar to those used in [3], [9] and [13] to find $\alpha$, $\mu$, $\rho$, $\tau$, $d_D$, $d_Q$, $d_t$, $d_b$, $d_c$, $E_n$, $\mu_n$ and others. We apply miss rate values obtained by [3] using PARSEC [5] and NAS [6] benchmarks. The average memory delay *vs.* total cache area under constrained area budget, off-chip memory bandwidth, power budget and NoC capacity are shown in Fig. 2 (a), (b), (c) and (d) respectively. The optimal area allocations per level presented as fraction of total cache area *vs.* area budget under the same constraints are shown in Fig. 3 (a), (b), (c) and (d) respectively. Under constrained area budget, as more area is allocated to cache, the hierarchy deepens and the average memory delay decreases to reach the optimum point (marked by rhombus shapes for each configuration in Fig. 2(a)); increasing area beyond that point leads to a longer memory delay, as discussed in Section 2. The entire area is initially allocated to $L_1$; when area becomes sufficient for two-level configuration, it is divided between $L_1$ and $L_2$, with $L_1$ fraction decreasing and $L_2$ growing with area; as area suffices for three-level configuration, it is divided among $L_1$, $L_2$ and $L_3$, with $L_3$ fraction growing at the expense of $L_1$ and $L_2$.

To satisfy the off-chip memory bandwidth constraint, two-level hierarchy is a minimal requirement, so that $M_D$ is sufficiently low; as area grows, three-level hierarchy becomes optimal at the area of ~40 (Fig. 2(b)); a single-level cache becomes a viable although suboptimal solution at ~110 (Fig. 2(b)). Accordingly, the $L_1$ area allocation is initially zero; as area grows sufficiently to enable two-level (at ~20, Fig. 3(b)), and then three level (at ~35, Fig. 3(b)) configurations, the allocation among levels continues similarly to the constrained area scenario.

The constrained power case resembles the constrained area scenario for area below ~140 (Fig. 2(c)); as area grows beyond that point, the power budget becomes insufficient for the three-level cache, so two-level hierarchy reverts to be optimal (at ~170, Fig. 2(c)); as area grows further and power budget remains limited, single level cache replaces the two-level configuration as the only viable solution (at ~240, Fig. 2(c)). The area allocation below ~140 is similar to the constrained area scenario; when $L_3$ is eliminated (at ~170, Fig. 3(c)), its area is divided between $L_1$ and $L_2$; as area continues to grow, $L_2$ grows at the expense of $L_1$; when $L_2$ in turn is removed (at ~240, Fig. 3(c)), the entire area budget is allocated to $L_1$.

Under the constrained NoC bandwidth, the single-level cache provides the optimal solution when area is low, although certain minimal area (~30, Fig. 2(d)) is required to keep $M_S$ in check; as area grows, the optimality point skips the two-level hierarchy to shift directly to the three level configuration (at ~35, Fig. 2(d)); as area grows further, two-level cache becomes viable although suboptimal (at ~70, Fig. 2(d)). Accordingly, the area budget is initially allocated to $L_1$ (~30, Fig. 3(d)), and then (at ~35) is divided among all levels, to further behave similarly to the constrained area or off-chip bandwidth scenarios.

## 4 CONCLUSIONS

This paper describes a cache hierarchy optimization framework that finds an optimal cache hierarchy and allocates hardware resources among hierarchy levels. The algorithm relies on modeling the cache access time as a function of cache size. The proposed framework allows performance optimization under typical CMP restrictions: constrained power budget, limited off-chip memory bandwidth, constrained area and limited NoC capacity.

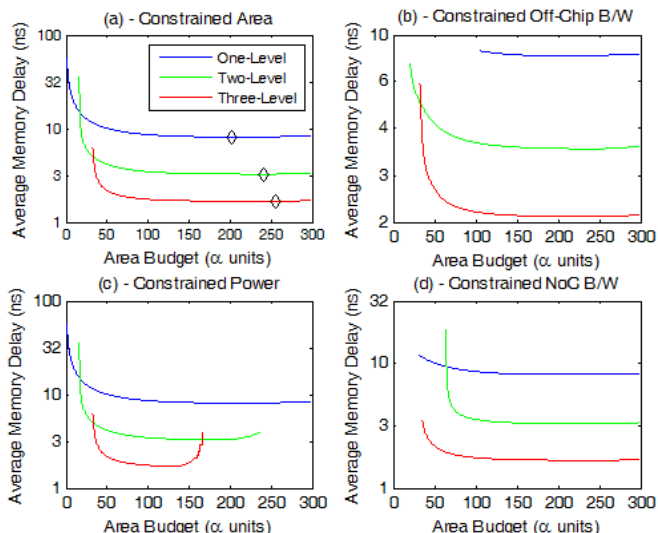

Fig. 2. Average Memory Delay *vs*. Area under: (a) constrained area, (b) constrained off-chip B/W, (c) constrained power, (d) constrained NoC B/W.

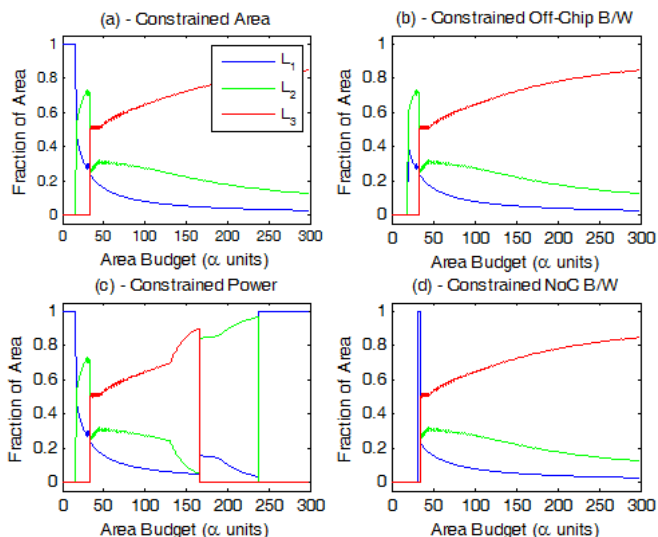

Fig. 3. Per-Level Fraction of Area *vs*. Area under; (a) constrained area, (b) constrained off-chip B/W, (c) constrained power, (d) constrained NoC B/W.

The optimization is extended by incorporating the impact of the data sharing on cache miss rate. We find that in power-constrained CMPs, area overcommitment shifts the optimal configuration from three levels back to two levels, and eventually back to a single level. In off-chip memory bandwidth-constrained CMPs, a single-level cache is not a practical solution hence a larger minimum area allocation is mandatory to support two- and three-level hierarchies. These results are in line with the findings of [10]. In NoC capacity-constrained CMPs, single-level might be preferable over two-level hierarchy; as area budget grows, the optimality point skips over two-level hierarchy directly to three-level configuration.

In real life, cache performance is affected by a combination of constrained chip resources. When cache area budget is limited, off-chip memory bandwidth is the number one constraint: the elevated miss rate of the shared cache causes the off-chip memory traffic to intensify. As a result, the average memory delay is affected by longer DRAM queuing delays. When cache area budget is substantial, power consumption becomes the primary constraint as cache turns out to be too large to power. The imbalanced area allocation between private and shared levels may increase the miss rate of the private cache and as a result cause higher congestion in the NoC and longer average memory delay.

We have provided the architect a practical analytical tool for cache hierarchy partitioning which leads to optimal memory access delay under constrained resources. This framework can be extended in a number of ways, for example it can be incorporated into the CACTI simulator to offer an optimal cache hierarchy.

## ACKNOWLEDGMENT

We thank Prof. Uri Weiser and Yaniv Ben Itzhak for their review and remarks. This research was partially funded by the ICRI-CI and Hasso-Plattner-Institut.